\documentclass{article}
\usepackage{amsmath,amssymb,epsfig,amsfonts,epsfig,graphicx}

\newtheorem{Theorem}{Theorem}
\newcommand{\cA}{{\cal A}}
\newcommand{\dx}{\,\mathrm{d}x}
\newcommand{\nn}{\nonumber}
\newcommand{\half}{\frac{1}{2}}
\newcommand{\ve}{\varepsilon}
\newcommand{\R}{\mathbb R}
\newcommand{\N}{\mathbb N}
\newcommand{\Z}{\mathbb Z}
\begin{document}

\title{Continuum limits of bistable spring models of carbon nanotube arrays
accounting for material damage}

\author{T. Blesgen\footnote{Max Planck Institute for Mathematics in the
Sciences, Inselstra\ss e 22, D-04103 Leip\-zig, Germany,
email: {\tt blesgen@mis.mpg.de}},
F. Fraternali\footnote{Department of Civil Engineering, University of
Salerno, 84084 Fisciano(SA), Italy,
email: {\tt f.fraternali@unisa.it}},
J.~R. Raney\footnote{Engineering and Applied Science, California Institute
of Technology, Pasadena, CA 91125, USA,
email: {\tt raney@caltech.edu}},
A. Amendola\footnote{Department of Civil Engineering, University of Salerno,
84084 Fisciano(SA), Italy,
email: {\tt adamendola@gmail.com}},
C. Daraio\footnote{Engineering and Applied Science, California Institute of
Technology, Pasadena, CA 91125, USA,
email: {\tt daraio@caltech.edu}}}
\date{\today}
\maketitle

\begin{abstract}
Using chains of bistable springs, a model is derived to investigate
the plastic behavior of carbon nanotube arrays with damage. We study the
preconditioning effect due to the loading history by computing analytically the
stress-strain pattern corresponding to a fatigue-type damage of the structure.
We identify the convergence of the discrete response to the limiting case of
infinitely many springs, both analytically in the framework of
Gamma-convergence, as well as numerically.
\end{abstract}

{\small
\vspace*{2mm}
\begin{center}
{\bf Keywords}
\end{center}
\vspace*{-2mm}
\hspace*{9mm}\begin{minipage}[t]{10.3cm}
Carbon nanotube arrays; bistable springs; multiscale behavior;
Mullins effect; permanent deformation.
\end{minipage}
} 

\vspace*{4mm}
\section{Introduction}
\label{secintro}
Because of their interesting combination of properties, including high
strength, low density, and high electrical and thermal conductivities, carbon
nanotubes (CNTs) have been of great interest as nanoscale elements in a variety
of applications, \cite{apps}. Aligned arrays of CNTs can be readily
synthesized to form foam-like materials that combine low density with a
desirable dissipative response, \cite{Cao2005,c31}.

There are a few interesting structural and mechanical features of these
materials: First, the thermal chemical vapor deposition process that is
typically used to synthesize the arrays \cite{Cao2005} results in a gradient
in physical properties (such as density) along the height of the structure.
This leads to strain localization during compression, with the majority of the
structure remaining undeformed while increasing strain results in the
sequential addition of highly localized buckles
\cite{yaglioglu2007,hut10}. Second, there is a large amount of strain
recovery after compression (typical of the CNT arrays that we study, which are
synthesized using a vapor phase catalyst, \cite{Cao2005}, but not for all
types of CNT arrays that are synthesized differently,
\cite{yaglioglu2007,bradford2011}). Third, the stress-strain response of the
material is hysteretic, with different loading and unloading paths. It has been
noted in the past that repeated compressive cycles result in a hysteresis of
decreasing area and decreasing stress at any given strain for the first few
cycles, \cite{Cao2005}. This effect, sometimes referred to as
\emph{preconditioning}, ceases after the initial few cycles, resulting in a
hysteresis of constant area for loading cycles thereafter. Finally, we observe
that this preconditioning effect is dependent on the maximum strain reached.
When the maximum strain of all previous loading cycles is exceeded, the stress
response obtained at these elevated strains is that of the un-preconditioned
material, as if it had never been compressed previously, \cite{Misra08}.

The features we have enumerated for the foam-like response of CNT arrays are
analogous to those observed in other soft materials such as filled rubbers.
In the context of rubbers, this response is associated with what is termed the
\emph{Mullins effect}, \cite{mullins1947,dorfmann2004}. Capturing these
features simultaneously in models has proven
difficult in the past for other materials, with frequent use of simplifying
assumptions that only allow models to match some of the experimental
observations (see, e.g., the \emph{idealized Mullins effect} modeled in
\cite{detommasi2006}).

We show in the present work that a suitable generalization of the mesoscopic
mass-spring model of CNT structures recently proposed by
\cite{FBAD10} is able to handle the material damage due to
preconditioning and permanent deformation within an effective one-dimensional
framework. We introduce  preconditioning-induced material damage by
setting to zero the stiffness of a suitable percentage of the bistable springs
that describe the response of the material at the microscopic scale.
This allows us to model permanent axial deformation of the
structure through the irreversible `annihilation' of the
springs with zero stiffness.

\section{Multiscale mass-spring models of CNT arrays}

\subsection{Bistable spring model at the microscopic scale}
\label{secmodel}

We model an infinitesimal portion of a CNT foam through the bistable spring
model described in \cite{FBAD10}, which we hereafter briefly
summarize. We assume that such a portion of the foam can be described as a
chain of $N+1$ lumped masses $m^0,\ldots, m^N$, with $m^0$ clamped at the
bottom of the chain. The adjacent masses are connected to each other through
bistable springs characterized by the axial strains
\begin{equation}
\label{eps1} 
\ve^i=\ve^i(u_N)=\frac{u_N^{i-1}-u_N^i}{h_N}, \ \ \ \ i=1,\ldots,N
\end{equation}
where $u_N^i$ is the axial displacement of the mass $m^i$ (positive upward),
$h_N:=L/N$ is the equal spacing between the masses, and
$u_N:=\{u_N^0,\ldots, u_N^N\}$. The potential $V^i$ and stress $\sigma^i$
vs strain $\ve^i$ laws of the generic spring are
\begin{align}
\label{V1}
V^i(\ve^i) &:= \left\{\! \begin{array}{ll}
V_a^i(\ve^i):=-k_0^i[\ve^i+\ln(1\!-\!\ve^i)], & \ve^i<\ve_a^i,\\
V_b^i(\ve^i):=c_1\!+\!\sigma_a^i\ve^i\!+\!\frac{1}{2}k_b^i(\ve^i\!-\!\ve_a^i)^2,
& \ve_a^i\le\ve^i\le\bar{\ve}_c^i,\\
V_c^i(\ve^i):=c_2\!-\!k_c^i[\ve^i\!-\!\ve_*^i\!+\!\ln(1\!-\!(\ve^i\!-\!\ve_*^i
))], & \bar{\ve}_c^i<\ve^i,
\end{array}\right.\\
\label{sigma1}
\sigma^i(\ve^i) &= {V^i}'(\ve^i)=\left\{\begin{array}{ll}
k_0^i\frac{\ve^i}{1-\ve^i},& \ve^i<\ve_a^i,\\
\sigma_a^i+k_b^i(\ve^i-\ve_a^i), & \ve_a^i\le\ve^i\le\bar{\ve}_c^i,\\
\frac{k_c^i(\ve^i-\ve_*^i)}{1-(\ve^i-\ve_*^i)},
& \bar{\ve}_c^i<\ve^i\end{array}\right.
\end{align}
where $k_0^i>0$, $k_b^i<0$, $k_c^i>0$, $\ve_a^i>0$ and
$\ve_c^i\ge\ve_a^i$ are constitutive parameters (five independent parameters);
the constants $c_1<0$ and $c_2>0$ are such that $V_a^i(\ve_a^i)=V_b^i(\ve_a^i)$,
$V_b^i(\bar{\ve}_c^i)=V_c^i(\bar{\ve}_c^i)$; and it results (compare with
Fig.~2 and Fig.~3 of \cite{FBAD10} for the notation)

\begin{align}
\label{epsstar}
\ve_*^i=\ve_c^i-\frac{\sigma_a^i}{k_c^i\!+\!\sigma_a^i},\quad
\bar{\ve}_c^i=\frac{\ve_c^i(k_c^i\!+\!\sigma_a^i)}{k_c^i\!+\!\sigma_c^i}+
\frac{(\sigma_c^i\!-\!\sigma_a^i)(k_c^i\!+\!\ve_c^ik_c^i\!+\!\ve_c^i\sigma_a^i)}
{(k_c^i\!+\!\sigma_a^i)(k_c^i\!+\!\sigma_c^i)}
\end{align}
with $\sigma_a^i=k_0^i\frac{\ve_a^i}{1-\ve_a^i}$, 
$\sigma_c^i=\sigma_a^i+k_b^i(\bar{\ve}_c^i-\ve_a^i)$.

\subsection{Plasticity and damage}
\label{sechyst}
At the microscopic scale, the bistable springs introduced above permit a
dynamic switching process between the phases (a) and (c), cf.
\cite{PT02,PT05}. As in \cite{PT05}, we name a response of the material
{\it plastic}, if the strain $\ve^i$ of a single spring exceeds the
threshold $\ve_a^i$. For a chain of $N$ springs, this can
be characterized by the occurrence of loading and unloading stress plateaux.

Within the current section, we rescale for simplicity $L$ to unity;
name (b) the unstable phase;
and regard a mesoscopic element of a CNT array as the limit
$N\rightarrow\infty$ of a series of $N$ microscopic springs.

Let $m$ denote the number of hysteresis cycles that have been applied to
the material during the previous loading history (up to different maximum
strains). We assume that such a loading path has severely weakened the
stiffness of $(1-\beta(m))N$ microscopic springs, for given $0<\beta(m)\le1$,
and that at the current time it holds for all $m\in\N$
\begin{align*}
(A1)\quad & k_c^i=k_0^i\quad\mbox{for all }i\in\N\\[-4pt]
(A2)\quad & k_0^1=k_0^2=\ldots=k_0^{\lfloor\beta N\rfloor}=k_0,\qquad
k_0^{\lfloor\beta N+1\rfloor}=\ldots=k_0^N=\delta\\[-4pt]
(A3)\quad & \ve^i_*=\ve_*\quad\mbox{for }i=1,\ldots,\beta N\\[-4pt]
(A4)\quad & \ve^i_a={\bar \ve}^i_c\quad\mbox{for }i=\beta N+1,\ldots, N.
\end{align*}
Condition~(A1) stipulates the {\it symmetry} of the microscopic springs. (A2)
states that the springs $\lfloor\beta N+1\rfloor$ to $N$ have stiffness
$k_0^i=\delta$, where $\delta>0$ is a small constant (\textit{damaged springs}).
We name \textit{undamaged} those springs with stiffness constant $k_0$
(springs $1$ to $\beta N$). 
For the present analysis we also require a certain smallness condition
on $\ve_a^i$ and $\overline{\ve}_c^i$ relating to strong pinning that
disappears for $N\to\infty$, see \cite{PT02}.

With $V^i$ given by (\ref{V1}), the mechanical energy of the structure is
\[ E_N(u_N):=\frac1N\,\sum_{i=1}^NV^i(\ve^i(u_N)). \]
Let $\sigma$ be the given total stress. The mesoscopic average strain is simply
$\ve(u_N):=\frac1N\sum_{i=1}^N\ve^i(u_N)$,
where $\ve^i$ denotes the strain associated with the $i$-th spring.
Following an original idea of \cite{PT05}, we model plasticity by the gradient
flow equations
\begin{equation}
\label{gradflow}
\nu\,\dot{\ve}^i(u_N)=-\frac{\partial\Phi_N}{\partial\ve^i}(\ve^1(u_N),\ldots,
\ve^N(u_N))
\end{equation}
with the total energy
\[ \Phi_N(\ve^1,\ldots,\ve^N):=\frac1N\sum_{i=1}^N\big[
V^i(\ve^i)-\sigma\ve^i\big]. \]
The evolution equation (\ref{gradflow}) lets $\ve^i$ evolve towards local
minimizers of $\Phi_N$.
We are interested in the limit $\nu\to0$ which amounts to infinitely fast
evolution such that $\ve(u_N)$ attains a local minimizer of $\Phi_N$.
First we construct the equilibrium points. Inside the $i$-th spring
element, the strain must satisfy the condition
\[ (V^i)'(\ve^i)=\sigma. \]
For given total stress $\sigma$, there are at most the three local minimizers
(using (A3))
\begin{subequations}
\label{eadef}
\begin{align}
\breve{\ve}_a^i(m) &= \frac{\sigma}{k_0^i+\sigma}, \qquad
\breve{\ve}_b^i(m)=\frac{\sigma-\sigma_a}{k_b^i}+\ve_a^i,\\
\breve{\ve}_c^i(m) &= \frac{\sigma(1+\ve_*)+k_0^i\ve_*}{k_0^i+\sigma}
=\breve{\ve}_a^i(m)+\ve_*.
\end{align}
\end{subequations}
Note that for the derivation of (\ref{eadef}), we require that $\delta$ is
positive.

In a loading or unloading experiment, the first spring located closer to the
bottom of the structure is the softest and yields first, changing its phase,
\cite{CNTIDENT}. Next, the second spring yields, and so forth, until the
$\beta N$-th spring. (Note that in accordance with (A4), the springs
$\beta N+1,\ldots, N$ with small spring constant $\delta$ do not flip.)
Therefore, similar to the case of $N$ identical springs, the total state of the
series of springs is still completely specified by two scalar parameters $p$
and $q$ and the additional parameter $\beta$. Here, $p$, $q$, $1-p-q$
denote the phase fractions of the minimizers $a$, $b$, and $c$, which
corresponds to having $\beta Np$, $\beta Nq$ and $\beta N(1-p-q)$ springs in
phase $a$, $b$, and $c$, respectively. We assumed here that $\beta Np\in\N$.

As $\ve\mapsto V^i(\ve)$ is concave in Regime~$b$ for all $i\in\N$,
if the elongation of a spring in the local minimum $\breve{\ve}_b^i$ is altered
by an arbitrarily small perturbation, it will move (according to the sign of
the perturbation) to either $\breve{\ve}_a^i$ or $\breve{\ve}_c^i$. As a
consequence, any system of $N$ springs with $q\not=0$ is unstable and we may
in the following restrict to the case $q=0$.

The average strain of a system with $\beta N$ springs in equilibrium and the
first $\beta Np$ springs in phase a fulfills the identity
\begin{align*}
\ve(m)
&= \frac1N\sum_{i=1}^{\beta N}\breve{\ve}_a^i(m)+\frac1N
\sum_{i=\beta Np+1}^{\beta N}\ve_*+\frac1N\sum_{i=\beta N+1}^N
\frac{\sigma}{\delta+\sigma}.
\end{align*}
Here we only study the limiting case $\delta\searrow0$ where there is no
further small correction of the damaged springs. In this case, we obtain
\begin{align}
\label{ss1}
\ve(m)
=\frac{\beta\sigma(m)}{k_0\!+\!\sigma(m)}\!+\!(1\!-\!\beta)
\!+\!\beta(1\!-\!p)\ve_*
=\frac{\sigma(m)\!+\!(1\!-\!\beta)k_0}{k_0+\sigma(m)}\!+\!\beta(1\!-\!p)\ve_*,
\end{align}
where we used (A1)--(A4), (\ref{sigma1}) and (\ref{eadef}); especially
$\breve{\ve}_c^i=\breve{\ve}_a^i+\ve_*$.

Resolving (\ref{ss1}), we get the stress-strain relationship for a $N$-springs
system
\begin{equation}
\label{ovsdef}
\sigma(\ve,m)=\frac{k_0(\ve-\ve_p+\beta-1)}{1-(\ve-\ve_p)}
\end{equation}
with $\ve_p(m):=\beta(1-p)\ve_*$.
The latter can in a natural way be identified with the plastic strain. From
(\ref{ovsdef}) we see that $\sigma$ only depends on $m$ and on the elastic
strain $\ve_{\mathrm{el}}:=\ve-\ve_p$.

\subsection{Analytic computation of the continuum limit}
\label{secgamma}
We identify the continuum limit of $E_N$ in the framework of
$\Gamma$-convergence (see, e.g., \cite{Braides2002}). This is not
a standard procedure as $V$ also depends on the spatial position.

Let $L=1$ (which can always be obtained by rescaling), and $\Omega:=(0,1)$.
For prescribed $l>0$, we impose the boundary conditions
\begin{equation}
\label{BC} u_0^N=0, u_N^N=l.
\end{equation}
This selection of boundary conditions appears natural for the discrete system
in the absence of next-to-nearest neighbor-interactions.

We now specify our assumptions on $V$. The mechanical pair-potential is a
function of the deformation gradient. In addition, in order to capture the
effect of damage expressed in the assumptions (A1)--(A4) above, we need to
respect the dependence of $V$ on the spatial position $i$. Let
$\mathrm{dom}(V)=[0,1]\times D$ for suitable $D\subset\R$ be the domain of
definition of $V$. We postulate the following conditions on $V$.

\vspace*{2mm}
(COND~1) There exist positive constants $c_1$, $C_1$, such that
\[ c_1(|\ve|-1)\le V(x,\ve)\le C_1(|\ve|+1)\quad\mbox{for all }
(x,\ve)\in[0,1]\times D. \]

(COND~2) The function $x\mapsto V(x,\ve)$ is continuous for any $\ve\in D$.
\vspace*{2mm}

\noindent After extending $V$ given by (\ref{V1}) continuously in $i$ by
interpolation, we easily verify that this extension (again called $V$)
satisfies both COND~1 and COND~2. For given deformations
$u_N:=(u_i^N)_{0\le i\le N}$, we may rewrite the overall mechanical energy of
the chain as a functional of the discrete displacement gradient,
\begin{equation}
\label{Edef}
E_N(u_N):=h_N\sum_{i=0}^{N-1} V\Big(\frac iN,\frac{u_{i+1}^N-u_i^N}{h_N}\Big).
\end{equation}
The functional $V$ coincides with (\ref{V1}), but the dependence on the
subscript $i$ has moved to the first argument. We scaled $E_N$ by $h_N$ as we
are dealing with a microscopic energy.

Following ideas in \cite{BDG}, we introduce for $N\in\N$ the set $\cA_N$ of
all functions $u: h_N\Z\cap[0,1]\to\R$, setting $u_i:=u(ih_N)$. We tacitly
identify $\cA(0,1)$ with the piecewise affine linear interpolations, i.e.
\[ \cA(0,1):=\big\{u:[0,1]\to\R\;\big|\;u \mbox{ is affine in }
(ih_N,(i+1)h_N),\,0\le i\le N-1\big\}. \]
With $u_i=u(ih_N)$, as a shorthand notation for the second argument in
(\ref{Edef}), we introduce the symbol
\[ \nabla^N u_N(x):=\frac{u_{i+1}^N-u_i^N}{h_N}, \quad
x\in [ih_N,(i+1)h_N),\;0\le i\le N-1, \]
which is a discrete approximation of $\nabla u$ with step size $h_N$.

For later use we introduce $E^l_N:\cA(0,1)\to\R\cup\{+\infty\}$ by
\[ E_N^l(u):=\left\{\!\!\begin{array}{ll} E_N(u), & \mbox{if }u(0)=0,\,u(1)=l,\\
+\infty, & \mbox{otherwise.} \end{array}\right. \]

\begin{Theorem}({\bf Continuum limit of $E_N$})
\label{theo1}\\
Let $V$ satisfy the conditions (COND~1), (COND~2) and let $l>0$. Then the
functional $E_N^l$ converges in the $\Gamma$-sense for $N\to\infty$ to a
functional $E^l: L^1(0,1)\to\R$ defined by
\[ E^l(u)\!\!:=\!\!\left\{\!\!\begin{array}{ll} 
\int_0^1 V_0^{**}(x,\nabla u(x))\dx &\mbox{if }u\in H^{1,1}(0,1),
u(0)=0,\,u(1)=l,\\
+\infty & \mbox{else.} \end{array}\right. \]
\end{Theorem}
Here, $V_0(x,z):=\half\min\Big\{V(x,z_1)+V(x,z_2)\;|\; z_1+z_2=2z\Big\}$,
and $V_0^{**}$ denotes the convexification of $V_0$
(see, e.g., \cite{Rockafellar}).

\vspace*{2mm}
\noindent{\bf Proof} (a) {\it Proof of the $\liminf$-inequality}.

Let a sequence $(u_N)_{N\in\N}\subset L^1(0,1)$ be given with $u_N\to u$ in
$L^1(0,1)$ as $N\to\infty$. W.l.o.g. $E_N^l(u_N)<\infty$ and
thus $E_N(u_N)<\infty$. Now, as the bounds in (COND~1) are uniform in $x$, we
can apply Theorem~2 in \cite{Blesgen07} which proofs the $\liminf$-inequality
for $(u_N)_N$.

\vspace*{2mm}
\noindent(b) {\it Proof of the $\limsup$-inequality}.

Let $u\in L^1(0,1)$ be given. We have to show the existence of a sequence
$(u_N)_{N\in\N}\subset L^1(0,1)$ such that $u_N\to u$ in $L^1(0,1)$ as
$N\to\infty$ and
\[ \limsup_{N\to\infty}E_N^l(u_N)\le E^l(u). \]
By a standard density argument, cf. Part~(b3) in the proof of Theorem~2 in
\cite{Blesgen07}, we may restrict to the affine case $u(x)=ax+b$.

For the construction we first ignore the boundary conditions (\ref{BC}).
Let $N=km$ for some $m,\,k\in\N$. We will choose functions $u_N$ which are
periodic in any subinterval of $(0,1)$ with length $mh_N$. We write the
integrand as $V(d_N(x),\cdot)$, where $d_N:(0,1)\to\R$ is a piecewise
constant bounded function with $d_N\to\mathrm{Id}$ for $N\to\infty$. The
method can be generalized to more general situations.
With these settings we find
\begin{align}
E_N(u_N)
\label{est1} =& \sum_{i=0}^{k-1}\int_{imh_N}^{(i+1)mh_N}
V(d^i,\nabla^N u_N(x))\dx+o(1),
\end{align}
where $d^i:=d_N((i+1/2)mh_N)$; and (COND~2) has been used.

By convexity of $V_0^{**}$ and Carath{\'e}odory's theorem
(see, e.g., \cite{Rockafellar})
we know that for any $0\le i\le k-1$, there exist real
numbers $\lambda_1$, $\lambda_2$ with $0\le\lambda_1,\,\lambda_2\le1$ and
$0\le\lambda_1+\lambda_2\le1$ such that
\begin{align}
\label{Cara}
V_0^{**}(d^i,a)=\lambda_1 V_0(d^{1,i},a^1)+\lambda_2 V_0(d^{2,i},a^2)
+(1-\lambda_1-\lambda_2)V_0(d^{3,i},a^3),\\
\label{aline}
(d^i,a)=(d^i,\nabla u)=\lambda_1(d^{i,1},a^1)
\!+\!\lambda_2(d^{i,2},a^2)\!+\!(1\!-\!\lambda_1\!-\!\lambda_2)(d^{i,3},a^3).
\end{align}
For given $\lambda_1$, $\lambda_2$ we introduce the sets
\begin{align*}
\Omega^N_1 &:= \Omega\cap\cup_{i=0}^{k-1}\;(ihm,
ihm+h\lfloor\lambda_1m\rfloor],\\
\Omega^N_2 &:= \Omega\cap\cup_{i=0}^{k-1}\;((ihm+h\lfloor\lambda_1m\rfloor),
(ihm+h\lfloor(\lambda_1+\lambda_2)m\rfloor],\\
\Omega^N_3 &:= \Omega\cap\cup_{i=0}^{k-1}\;(
(ihm+h\lfloor(\lambda_1+\lambda_2)m\rfloor),(i+1)hm],
\end{align*}
such that $\Omega=\Omega^N_1\cup\Omega^N_2\cup\Omega^N_3$.

By definition of $V_0$ it holds for $0\le i\le k-1$ and any $s=1,2,3$,
\[ V_0(d^{s,i},a^s)=\half\Big[V(d_1^{s,i},a^s_1)+V(d_2^{s,i},a^s_2)\Big], \]
where $d^{s,i}$, $a^s$, $1\le s\le3$, are suitable numbers such that
$d^{s,i}=\frac{d^{s,i}_1+d^{s,i}_2}{2}$ and $a^s=\frac{a^s_1+a^s_2}{2}$.
We choose $u_N(x)=a_N(x)x+b$, where
\[ a_N(x)\!=\!\!\left\{\!\! \begin{array}{ll}
a_1^1, &\mbox{if}\;x\in\Omega^N_1\cap\cup_{i=0}^{k-1}\cup_{j=0}^{m/2-1}\,
((im\!+\!2j)h_N,im\!+\!2j\!+\!1)h_N],\\
a_1^2, &\mbox{if}\;x\in\Omega^N_2\cap\cup_{i=0}^{k-1}\cup_{j=0}^{m/2-1}\,
((im\!+\!2j)h_N,(im\!+\!2j\!+\!1)h_N],\\
a_1^3, &\mbox{if}\;x\in\Omega^N_3\cap\cup_{i=0}^{k-1}\cup_{j=0}^{m/2-1}\,
((im\!+\!2j)h_N,(im\!+\!2j\!+\!1)h_N],\\
a_2^1, &\mbox{if}\;x\in\Omega^N_1\cap\cup_{i=0}^{k-1}\cup_{j=0}^{m/2-1}\,
((im\!+\!2j\!+\!1)h_N,(im\!+\!2j\!+\!2)h_N],\\
a_2^2, &\mbox{if}\;x\in\Omega^N_2\cap\cup_{i=0}^{k-1}\cup_{j=0}^{m/2-1}\,
((im\!+\!2j\!+\!1)h_N,(im\!+\!2j\!+\!2)h_N],\\
a_2^3, &\mbox{if}\;x\in\Omega^N_3\cap\cup_{i=0}^{k-1}\cup_{j=0}^{m/2-1}\,
((im\!+\!2j\!+\!1)h_N,(im\!+\!2j\!+\!2)h_N].\end{array}\right. \]
For $s=1,2,3$ and $0\le i\le k-1$ we set
\[ d_N(x)=\left\{\! \begin{array}{ll}
d_1^{s,i}, &\mbox{if }x\in\Omega^N_s\cap\cup_{j=0}^{m/2-1}\;
((im+2j)h_N,(im+2j+1)h_N],\\
d_2^{s,i}, &\mbox{if }x\in\Omega^N_s\cap\cup_{j=0}^{m/2-1}\;
((im+2j+1)h_N,(im+2j+2)h_N].\end{array}\right. \]
The ansatz for $d_N$ depends on the current interval, as $V$ is $x$-dependent.

Eqn.~(\ref{est1}) now reads after setting
$\lfloor\lambda_3m\rfloor\!:=\!m\!-\!\lfloor\lambda_1m\rfloor\!-\!\lfloor
\lambda_2m\rfloor$ for short
\begin{align}
E_N(u_N) &= h_N\sum_{i=0}^{k-1}\sum_{s=1}^3\lfloor\lambda_sm\rfloor\;\half
\Big[V(d_1^{s,i},a_1^s)+V(d_2^{s,i},a_2^s)\Big]\nn\\
\label{conx}
&= \sum_{i=0}^{k-1}\hspace{-1.45em}-\hspace{0.45em}\sum_{s=1}^3
\frac{\lfloor\lambda_sm\rfloor}{m}V_0(d^{s,i},a^s).
\end{align}
For $m\to\infty$ we have $\lfloor\lambda_s m\rfloor/m\to\lambda_s$, $s=1,2$.
Consequently, using (\ref{Cara}),
\[ E_N(u_N)\to\int_0^1 V_0^{**}(x,\nabla u(x))\dx=E^l(u)\quad\mbox{as }
N\to\infty. \]
We still have to show that $u_N\to u$ in $L^1(0,1)$.
If in the derivation of Eqn.~(\ref{conx}) we formally set $V(x,v):\equiv v$
(which is feasible), we obtain
\[ \int_0^1\nabla u_N(x)\dx=\frac{\lfloor\lambda_1 m\rfloor}{m}a^1
+\frac{\lfloor\lambda_2m\rfloor}{m}a^2
+\Big(1-\frac{\lfloor\lambda_1m\rfloor}{m}-\frac{\lfloor\lambda_2m\rfloor}{m}
\Big)a^3 \]
and in the limit $m\to\infty$ as above
\begin{equation}
\label{con}
\lim_{N\to\infty} \int_0^1\nabla u_N(x)\dx=\lambda_1 a^1
\!+\!\lambda_2a^2\!+\!(1\!-\!\lambda_1\!-\!\lambda_2)a^3=a=
\int_0^1\nabla u(x)\dx,
\end{equation}
where Eqn.~(\ref{aline}) has been used. By (\ref{con}), $u_N\to u$ in
$L^r(0,1)$ for $1\le r\le\infty$.

\vspace*{2mm}
We still need to incorporate the boundary condition (\ref{BC}).
If $u_N$ is the recovery sequence from above, we define
\[ v_N(x):=\left\{\!\!
\begin{array}{rl} \frac{u_N(h_N)}{h_N}x, & x\in[0,h_N),\\
u_N(x), & x\in[h_N,1-h_N],\\
\frac{l-u_N(1-h_N)}{h_N}(x-1)+l, & x\in(1-h_N,1]. \end{array}\right. \]
By construction, $v_N$ is continuous on $[0,1]$ with $v_N(0)=0$, $v_N(1)=l$, and
$\lim_{N\to\infty}(E_N(u_N)-E_N(v_N))=0$,
which shows that $v_N$ is the sought recovery sequence.$\qquad\Box$

\section{Numerical results}
\label{secnum}
We study in this section
the numerical convergence of the stress-strain response of finite
mass-spring systems to the continuum limit of Eqn.~(\ref{ovsdef}).
We analyze the overall loading-unloading response of discrete systems composed
of $N=x+y$ springs, where $x$ denotes the number of undamaged springs,
while $y$ specifies the number of damaged springs.
For the spring constants, we use the parameters $k_0=50.00\times10^6$ Pa,
$k_b=-22.44\times10^6$ Pa, $\ve_a=0.25$, $\ve_*=0.52$,
$\delta=50.00\times10^2$ Pa, which correspond to $\sigma_a=16.67$ MPa,
$\sigma_c=5.00$ MPa, and $\Delta\sigma=\sigma_c-\sigma_a=-11.67$ MPa.
\begin{figure}[hb]
\unitlength1cm
\begin{picture}(12.0,7.8)
\put(-0.3,3.65){\psfig{figure=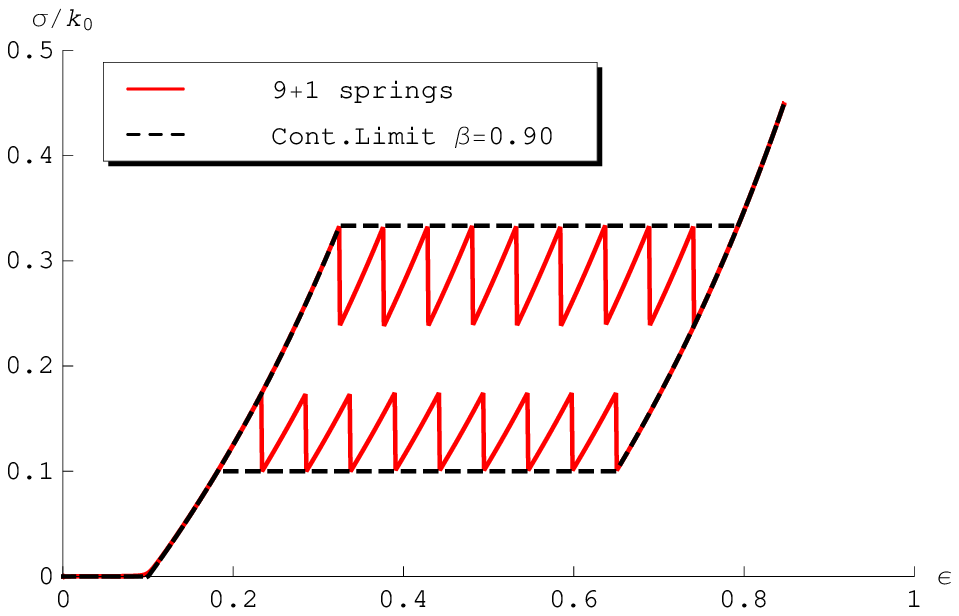,height=4.5cm}}
\put(6.3,3.65){\psfig{figure=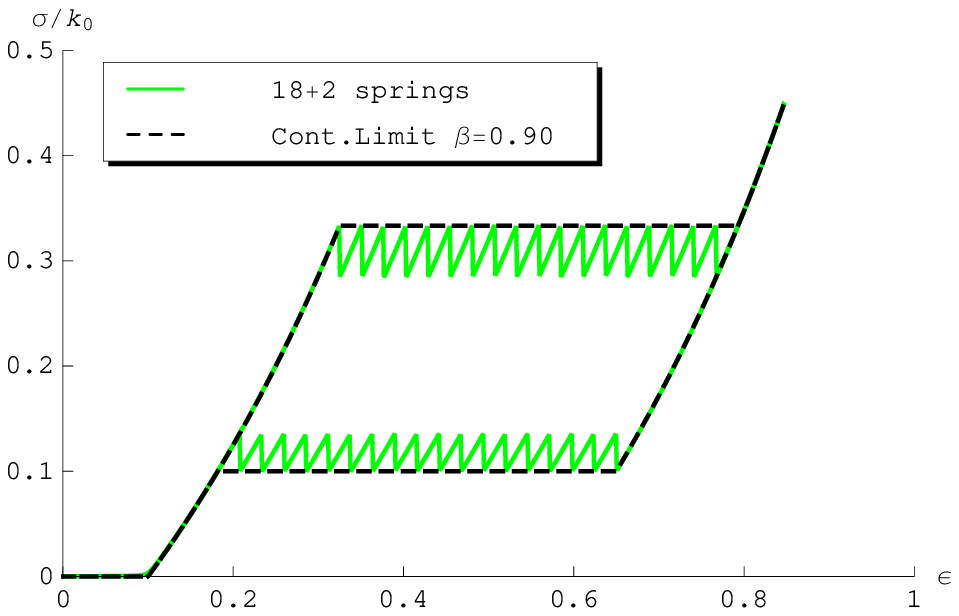,height=4.5cm}}
\put(-0.3,-0.55){\psfig{figure=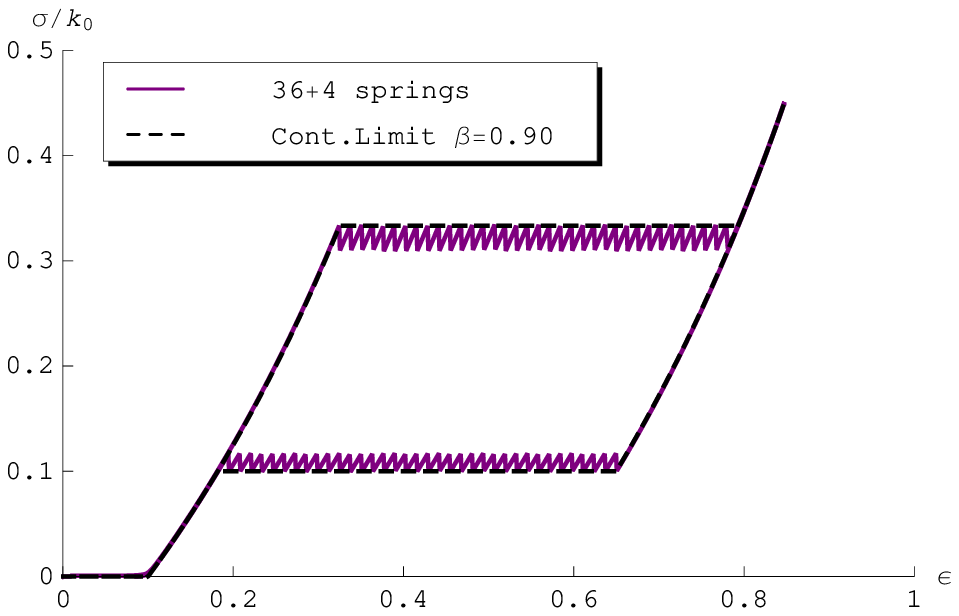,height=4.5cm}}
\put(6.3,-0.55){\psfig{figure=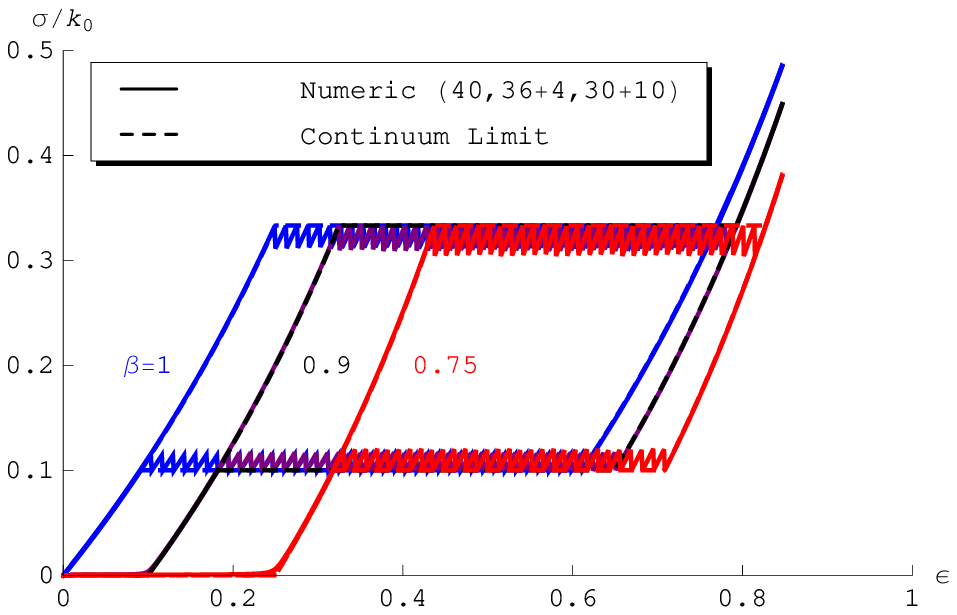,height=4.5cm}}
\end{picture}
\caption{Numerical overall stress-strain response of bistable mass-spring
chains for different total number of springs $N$.}
\label{fig:convergence_2011}
\end{figure}

Fig.~\ref{fig:convergence_2011} shows the results for different $x$, $y$.
The stress-strain curves of the discrete microscopic chains follow
sawtooth patterns which alternate \textit{elastic} and \textit{plastic}
steps. The plastic steps are characterized by fixed total strain and
microscopic branch switching.
Such sawtooth-like responses converge to a perfectly plastic behavior with
a loading ($\sigma=\sigma_a$) and an unloading plateau ($\sigma=\sigma_c$) for
increasing values of $N$, as predicted by
Eqn.~(\ref{ovsdef}). It is worth observing that the stress is zero for
$\ve\le1-\beta$, while for $\ve>1-\beta$ that the system is able to bear
stresses $\sigma>0$. Fig.~\ref{fig:convergence_2011} shows, in
addition, that the energy dissipation capacity of the system (area enclosed by
stress-strain curve in a loading-unloading cycle) reduces for decreasing
$\beta$, that is for increasing number of damaged springs (bottom-right
panel). Remarkably, the quantity $1-\beta$ can be regarded as the `activation'
strain of the system.

\section{Concluding remarks}
\label{secdiscussion}
In this paper, for the first time a bistable-spring ansatz has been proposed
capable of incorporating damage in plastically deformed materials.

We have shown that a suitable modification of the model recently proposed by
\cite{FBAD10} for CNT foams is able to handle preconditioning induced material
damage, which is characterized by an activation strain different from zero;
a reduction in the energy dissipation capacity; and permanent deformation. The
latter, in particular, coincides with the activation strain.
The new model allows us to extend the `transformational plasticity' concept
discussed by \cite{PT05} from time-independent hysteretic behavior to
fatigue-type material damage. 
It applies to a wide class of materials showing Mullins-like behavior,
\cite{mullins1947}, which includes besides CNT arrays
rubber-like and soft biological materials.

We address a multiscale formulation of the present model, accounting for graded
mechanical properties along the height of the structure, \cite{CNTIDENT}, and
the analytic computation of the energy dissipated by the system in the
continuum limit in future work.

\section*{Acknowledgements}
FF acknowledges the support of the Italian Network of Seismic Engineering
Laboratories (ReLUIS) and the Italian Civil Protection Department (DPC),
through the ReLUIS-DPC grant 2010/2013. TB acknowledges the support of the
German Research Community (DFG) through grant BL 512 4/1. JRR gratefully
acknowledges the U.S. Department of Defense and the Army Research Office for
their support via a National Defense Science \& Engineering Graduate (NDSEG)
fellowship. CD acknowledges support from the Institute for Collaborative
Biotechnologies under contract W911NF-09-D-0001 with the Army Research Office.

\bibliographystyle{elsarticle-harv}

\end{document}